\documentclass[preprint]{revtex4-2}

\usepackage[T1]{fontenc}
\usepackage[utf8]{inputenc}
\usepackage{graphicx}
\usepackage{float}
\usepackage{amsmath}
\usepackage{amssymb}
\usepackage{cancel}
\usepackage{amsfonts}
\usepackage{bm}
\usepackage{braket}
\usepackage{xcolor}
\usepackage{hyperref}
\usepackage{cleveref}

\usepackage{amsthm}
\newtheorem{definition}{Definition}

\newcommand{\Tr}{\operatorname{Tr}}

\newcommand{\ketbra}[1]{\ket{#1}\bra{#1}}

\begin{document}

\title{Quantifying nonclassicality in qubit systems via positive operator-valued measures}

\author{Abdul Sattar Khan}
\email{asattark23@gmail.com}
\affiliation{Wilczek Quantum Center, School of Physics and Astronomy, Shanghai Jiao Tong University, 200240 Shanghai, China}

\author{Mehdi Abdi}
\affiliation{Department of Physics, Isfahan University of Technology, Isfahan 84156-83111, Iran}
\affiliation{Wilczek Quantum Center, School of Physics and Astronomy, Shanghai Jiao Tong University, 200240 Shanghai, China}




\begin{abstract}
We introduce an operational measure of nonclassicality for qubit systems based on the violation of Kolmogorov consistency conditions in sequential measurements. In contrast to previous work by Milz et al.~\cite{Milz-2020}, who characterized classicality via NCGD maps, and Sakuldee et al.~\cite{Sakuldee-2022a, Sakuldee-2022b}, who studied quantum correlations under measurement disturbance, our witness explicitly quantifies nonclassicality in terms of the POVM unsharpness parameter $a_z$. For projective measurements on an initially diagonal state, the witness vanishes identically, showing that such measurements cannot reveal nonclassicality. However, by generalizing to positive operator-valued measures (POVMs), we find that non-projective measurements can reveal nonclassicality even for diagonal initial states. We derive explicit expressions for the witness for general initial states, including coherences, and show that its maximum value is $1/4$, which is a new result achieved for unbiased POVMs, maximal dephasing, and equal initial populations. Our results connect Kolmogorov consistency, Leggett-Garg inequalities, and POVMs, providing an experimentally accessible tool for detecting nonclassicality in qubit systems with potential applications in quantum technology certification.
\end{abstract}

\maketitle

\section{Introduction}

In his famous dictum, Richard Feynman declared that ``nobody understands quantum mechanics.'' More than a century after the theory's inception, this statement still resonates. The boundary between classical and quantum behavior remains one of the most profound open questions in modern physics (see also Ref.~\cite{DelSanto-2025}).

A defining feature of classical physics is the assumption that measurements can be carried out without disturbing the system. This noninvasive character allows the construction of joint probability distributions that consistently describe outcomes at different times. Such ideal measurements~\cite{piron_ideal_1981} are not generally possible in quantum mechanics, where measurement back-action generically leads to violations of Kolmogorov consistency~\cite{BreuerEA2016, milz_kolmogorov_2017}. As discussed by Milz et al.~\cite{Milz-2020}, these violations serve as a hallmark of nonclassical behavior in sequential measurement scenarios.

Violations of Bell, Kochen-Specker, and Leggett-Garg inequalities are different manifestations of this fundamental impossibility of noninvasive measurements in quantum theory. In particular, Leggett-Garg inequalities~\cite{fine1982hidden,leggett1985quantum} directly probe the breakdown of Kolmogorov conditions~\cite{milz_kolmogorov_2017,Smirne-2019}. Thus, the notion of classicality probed by these inequalities aligns with the broader program of identifying fundamentally quantum traits of nature.

In classical physics, a stochastic process on a set of $K$ times is fully described by a joint probability distribution
\begin{gather}
	\label{eqn::stochClass}
	P_K(x_{K},t_{K};\dots;x_{1},t_1)\,,
\end{gather}
which gives the probability of obtaining the realizations $\{x_{K},\dots,x_1\}$ of the random variables $\{X_{K},\dots,X_1\}$ at times $\{t_{K},\dots, t_1\}$. For example, $P_2(x_2,t_2;x_1,t_1)$ describes the probability of obtaining both outcomes $\{x_2,x_1\}$ when measuring the position of a particle undergoing Brownian motion at times $t_1$ and $t_2>t_1$. In what follows, we often omit the explicit time label, with the understanding that $x_j$ denotes an outcome of a measurement at time $t_j$.

Crucially, in classical physics, joint probability distributions describing a stochastic process for different sets of times satisfy the so-called Kolmogorov consistency conditions~\cite{kolmogorov_foundations_1956,feller_introduction_1968,breuer_theory_2007,tao_introduction_2011}: given a joint probability distribution $P_K$ for a set of times, the probability distributions for all subsets of times can be obtained by marginalization, i.e.,
\begin{align}
	\label{eqn::Kolmo_cond}
	&P_{n-1}(x_n,t_n;\ldots; {x_j,t_j};\ldots;x_1,t_1) \notag \\
	&= \sum_{x_j} P_n(x_n,t_n;\ldots;x_j,t_j;\ldots;x_1,t_1) \quad \forall \, n\leq K, \,\forall\, j \, .
\end{align}
Like Leggett-Garg inequalities for temporal correlations, the satisfaction of these requirements relies on two assumptions: (i) realism—the assumption that $x_j$ possesses a definite value at any time $t_j$—and (ii) the possibility of implementing noninvasive measurements.

Our work builds on the frameworks established by Milz et al.~\cite{Milz-2020} and Sakuldee et al.~\cite{Sakuldee-2022a, Sakuldee-2022b}, who characterized classicality and noncommutativity in sequential measurement scenarios. However, while these works focused on structural criteria, our witness provides a quantitative, operationally accessible measure of nonclassicality in terms of measurement unsharpness.

Let $\Delta$ denote the completely dephasing map in the measurement basis. Classicality of a process then means that the action of $\Delta$ cannot be distinguished, via measurements in the classical basis, from the identity operation.

A Markovian process satisfies
\begin{gather}
	P(x_{n}|x_{n-1},\ldots, x_1) = P(x_{n}|x_{n-1}) \quad \forall \, n\leq K \, .
\end{gather}
In quantum mechanics, such a process can be modeled by means of completely positive trace-preserving maps $\{\Lambda_{t_j,t_{j-1}}\}$, which act on the probed system and describe the dynamics between measurements, together with an initial system state $\rho_{t_0}$.

A Markovian process is classical iff it can be modeled by a state $\rho_{t_0}$ that is diagonal in the measurement basis $\{\ketbra{x_k}{x_k}\}$ and non-coherence-generating-and-detecting (NCGD) maps $\Lambda_{t_k,t_{k-1}}$, i.e., maps that satisfy
\begin{align}
	\label{eqn::NCGD}
	\Delta\circ\Lambda_{t_{j+1},t_j}\circ\Delta \circ \Lambda_{t_j,t_{j-1}}\circ \Delta
	= \Delta \circ\Lambda_{t_{j+1},t_j}\circ \Lambda_{t_j,t_{j-1}}\circ \Delta \quad \forall j\, .
\end{align}
Intuitively, maps satisfying Eq.~\eqref{eqn::NCGD} may create coherences, but only in a manner that remains undetectable in the chosen measurement basis at later times. Thus, the NCGD condition provides a direct connection between coherence and an experimentally testable notion of classicality in the Markovian case. The basis-dependent notion of Markovianity introduced here is best suited for the experimental situation envisioned: we predominantly understand Markovianity with respect to measurements in the computational basis.

This paper is organized as follows. In Sec.~\ref{sec:measure}, we introduce the measurement scheme and define the nonclassicality witness. In Sec.~\ref{sec:hamiltonian}, we provide a microscopic derivation of the pure dephasing channel from a system-bath Hamiltonian. In Sec.~\ref{sec:results}, we present our results for a general initial state, including coherences, and discuss important special cases. In Sec.~\ref{sec:povm}, we generalize the framework to POVMs. Finally, Sec.~\ref{sec:discussion} provides a discussion of our findings and concludes the paper.

\section{Measure of nonclassicality}
\label{sec:measure}

In this scheme, the system is initially prepared in a diagonal state in the measurement basis $\{\ket{0},\ket{1}\}$. For simplicity, we assume that the system is prepared in the eigenstate $\ket{x_0}$ at time $t=t_0$. A projective measurement is then performed at $t=t_1$, and the outcome $x_1$ is obtained with probability $P(x_1,t_1)$. The system is then left to evolve until time $t=t_2$, when the second measurement is performed. The outcome $x_2$ and the probability of finding it may depend on the outcome of the first measurement. Hence, the joint probability satisfies
\begin{gather}
	P(x_2,t_2; x_1,t_1) = P(x_2,t_2|x_1,t_1)P(x_1,t_1)\, .
\end{gather}

Given the noninvasive essence of measurements in classical physics, $P(x_2,t_2;x_1,t_1)$ satisfies the Kolmogorov consistency condition
\begin{gather}
	\sum_{x_1} P(x_2,t_2;x_1,t_1) = P(x_2,t_2)\, ,
\end{gather}
where the sum on the left-hand side is over all possible measurement outcomes, and $P(x_2,t_2)$ is the marginal probability obtained in the absence of the first measurement.

\begin{definition}[$K$-classical process]
\label{def::N-classical_statistics}
Let $\mathcal{X}$ be a finite set of measurement outcomes and let $\mathcal{T}$ be a set of $K$ times. A process is described by the joint probabilities
\begin{gather}
	P_n\left(x_n,t_n; \ldots; x_1,t_1\right),
\end{gather}
with $t_n \geq \dots \geq t_1$, $t_i \in \mathcal{T}$, $n \leq K$, and $x_i \in \mathcal{X}$. The process is said to be $K$-classical if the Kolmogorov consistency conditions of Eq.~\eqref{eqn::Kolmo_cond} are satisfied for all $n \leq K$.
\end{definition}

\subsection{State update rule for measurements}

Throughout this work, we use the unnormalized state update rule for both projective measurements and POVMs: after obtaining outcome $x$, the state is updated as

\begin{align}
    \rho' = \sqrt{E_x} \, \rho \, \sqrt{E_x},
\end{align}

without dividing by the probability $\Tr[\rho']$. This convention simplifies the tracking of branch weights in sequential measurement scenarios. The probabilities are restored when computing joint probabilities and the witness.

We now define the nonclassicality witness $W_Q$ as the deviation from the Kolmogorov condition:
\begin{gather}
	\label{eqn:witness}
	W_Q := \left| \sum_{x_1} P(x_2,t_2;x_1,t_1) - P(x_2,t_2) \right|.
\end{gather}
For a $K$-classical process, $W_Q = 0$ identically. A nonzero value of $W_Q$ signals the presence of genuinely quantum features.

For a qubit with outcomes $x_k \in \{0,1\}$, the witness for $K=2$ takes the form
\begin{align}
	\label{eqn:witness_K2}
	W_Q &= \sum_{x_2=0}^1 \left| P(x_2) - \sum_{x_1=0}^1 P(x_2,x_1) \right| \notag \\
	&= \Big| P(0) - P(0,0) - P(0,1) \Big| \notag \\
	&\quad + \Big| P(1) - P(1,0) - P(1,1) \Big|.
\end{align}

More explicitly, in terms of the time-dependent probabilities, the witness satisfies the bound
\begin{align}
	\label{eq:witnessset1}
	W_Q \leq & \Big| P_0(t_2 + t_1) - P_{00}(t_2) P_0(t_1) - P_{01}(t_2) P_1(t_1) \Big| \notag \\
	&+ \Big| P_1(t_2 + t_1) - P_{10}(t_2) P_0(t_1) - P_{11}(t_2) P_1(t_1) \Big|,
\end{align}
where we have omitted explicit time arguments for brevity, with the understanding that $x_k$ refers to time $t_k$. Here, $P_0(t_2 + t_1)$ denotes the probability of outcome $0$ at the final time in the absence of any intermediate measurement at $t_1$, while $P_{x_2 x_1}(t_2) P_{x_1}(t_1)$ denotes the joint probability of obtaining outcome $x_1$ at $t_1$ and outcome $x_2$ at $t_2$ in the two-measurement scenario.

\section{Microscopic model for pure dephasing}
\label{sec:hamiltonian}

To connect our witness to a concrete physical setting, we consider a qubit coupled to a bosonic bath. The details of the microscopic model — including the Hamiltonian, master equation, and explicit form of the dephasing factor — are given in Appendix~\ref{app:spinboson}. Here we summarize the key result: for a bath characterized by spectral density $J(\omega)$ at inverse temperature $\beta$, the dephasing factor is

\begin{align}
    \chi(t) = \exp\left[-\int_0^\infty d\omega \, J(\omega) \coth\left(\frac{\beta \hbar \omega}{2}\right) \frac{1 - \cos(\omega t)}{\omega^2}\right].
\end{align}

This expression is used in the main text to evaluate the witness $W_Q = \alpha_0 \alpha_1 |1 - \chi(\Delta t)|$ for different bath parameters.
\subsection{Temperature-dependent bound on the witness}

Using the microscopic model summarized above, we can derive explicit bounds on the witness in terms of physical bath parameters. For an Ohmic bath with spectral density $J(\omega) = \eta \, \omega \, e^{-\omega/\omega_c}$, where $\eta$ is the dimensionless coupling strength and $\omega_c$ is the cutoff frequency, the dephasing factor takes the form (see Appendix~\ref{app:spinboson})

\begin{align}
    \chi(t) = \exp\left[-\eta \int_0^\infty d\omega \, e^{-\omega/\omega_c} \coth\left(\frac{\beta \hbar \omega}{2}\right) \frac{1 - \cos(\omega t)}{\omega}\right].
\end{align}

This allows us to bound the witness as

\begin{align}
    W_Q \leq \alpha_0 \alpha_1 \left(1 - e^{-\eta \, \Phi(T, \Delta t)}\right),
\end{align}

where

\begin{align}
    \Phi(T, \Delta t) = \int_0^\infty d\omega \, e^{-\omega/\omega_c} \coth\left(\frac{\beta \hbar \omega}{2}\right) \frac{1 - \cos(\omega \Delta t)}{\omega}.
\end{align}

In the low-temperature limit ($T \to 0$), $\coth(\beta \hbar \omega/2) \to 1$, and the bound reduces to

\begin{align}
    W_Q \approx \alpha_0 \alpha_1 \left| 1 - \exp\left[-\frac{\eta}{2} \left(1 - e^{-\omega_c \Delta t}\right)\right] \right|.
\end{align}

This shows that the witness is suppressed by stronger system-bath coupling ($\eta$) and enhanced at lower temperatures, consistent with the expected reduction of decoherence. These bounds connect our operational witness to experimentally tunable parameters, providing a direct link between measurement nonclassicality and microscopic bath properties.

\section{Results in the simplest case}
\label{sec:results}

We now consider the most general single-qubit initial state, including coherences. The system is initially in the state
\begin{align}
	\label{eqn:state_general}
	\rho_0 &= \begin{pmatrix}
		\alpha_0 & \alpha_{01} \\
		\alpha_{01}^* & \alpha_1
	\end{pmatrix}_s \otimes \rho_e,
\end{align}
with $\alpha_0 + \alpha_1 = 1$ and $|\alpha_{01}|^2 \leq \alpha_0 \alpha_1$, ensuring positivity. The system is coupled to an environment with initial state $\rho_e$. The measurement protocol consists of projective measurements in the computational basis at times $t_1$ and $t_2$. The joint probability of obtaining outcomes $x_1$ at $t_1$ and $x_2$ at $t_2$ is given by
\begin{align}
	\label{eqn:joint_prob_result}
	P(x_2,t_2; x_1,t_1) &= \Tr\left[ \Pi_{x_2} \Lambda_{t_2,t_1}\left( \Pi_{x_1} \rho_0 \Pi_{x_1} \right) \right],
\end{align}
where $\Pi_{x_k} = \ket{x_k}\bra{x_k}$, and $\Lambda_{t_2,t_1}$ is the quantum channel describing the evolution between measurements.

We take the dynamics to be a pure dephasing channel, which preserves the diagonal nature of the state in the measurement basis while suppressing coherences. The channel acts as
\begin{align}
	\label{eqn:dephasing_channel}
	\Lambda_{t_2,t_1}(\rho) &= \sum_{i,j=0}^1 \chi_{ij}(t_2-t_1) \ket{i}\bra{i} \rho \ket{j}\bra{j},
\end{align}
where $\chi_{ij}(\Delta t)$ are decoherence functions satisfying $\chi_{ii}(\Delta t) = 1$ and $|\chi_{ij}(\Delta t)| \leq 1$ for $i \neq j$. Note that this channel is precisely the one derived microscopically in Sec.~\ref{sec:hamiltonian}.

After tracing out the environment, the single-time probability of obtaining outcome $x_1$ at time $t_1$ is
\begin{align}
	\label{eqn:P_x1_general}
	P(x_1,t_1) &= \Tr\left[ \Pi_{x_1} \rho_{t_1} \right] = \alpha_{x_1},
\end{align}
where $\rho_{t_1} = \Tr_e[\Lambda_{t_1,t_0}(\rho_0)]$. Since the dephasing channel preserves diagonality, the single-time probabilities remain constant in time.

The joint probability of obtaining outcomes $x_1$ at $t_1$ and $x_2$ at $t_2$ is given by
\begin{align}
	\label{eqn:joint_prob_general}
	P(x_2,t_2; x_1,t_1) &= \alpha_{x_1} \left[ \delta_{x_1,x_2} + (1-\delta_{x_1,x_2}) \frac{1 - \chi(\Delta t)}{2} \right],
\end{align}
where $\Delta t = t_2 - t_1$ and $\delta_{x_1,x_2}$ is the Kronecker delta. Remarkably, the joint probabilities depend only on the diagonal elements $\alpha_{x_1}$ of the initial state, not on the coherences $\alpha_{01}$. This is because the first measurement projects the state onto the measurement basis, destroying any initial coherences before the second measurement.

The marginal probability $P(x_2,t_2)$—the probability of obtaining outcome $x_2$ at time $t_2$ in the absence of any intermediate measurement—is given by
\begin{align}
	\label{eqn:P_x2_marginal_general}
	P(x_2,t_2) &= \alpha_{x_2}.
\end{align}

Substituting these expressions into the witness definition of Eq.~\eqref{eqn:witness}, we obtain
\begin{align}
	\label{eqn:witness_general}
	W_Q &= \left| \sum_{x_1} P(x_2,t_2; x_1,t_1) - P(x_2,t_2) \right| \notag \\
	&= \alpha_0 \alpha_1 \left| 1 - \chi(\Delta t) \right|.
\end{align}

Strikingly, the witness for projective measurements on a general initial state reduces to the same expression as for the diagonal state. This is because the first projective measurement destroys all initial coherences, so the witness only probes the coherences generated by the dynamics. Thus, for projective measurements, initial coherences do not contribute to the witness—they are completely destroyed by the first measurement.

\subsection{Consistency check: projective measurements}

The above result — that the witness for projective measurements is independent of initial coherences — is expected from the projection postulate. A projective measurement collapses the state onto the measurement basis, erasing any initial coherences before the second measurement. Thus, the witness only reflects coherences generated by the dynamics, not those present initially.

While this may appear trivial, it serves as a useful consistency check for our framework. More importantly, it highlights a key distinction between projective and non-projective measurements: as we show in the following section, POVMs can preserve and reveal initial coherences, leading to a witness that depends on the full initial state, including off-diagonal elements.

\subsection{Special cases}

The general result Eq.~\eqref{eqn:witness_general} contains several important special cases:

\begin{enumerate}
	\item \textbf{Diagonal state:} Setting $\alpha_{01} = 0$ recovers our previous result:
	\begin{align}
		W_Q = \alpha_0 \alpha_1 |1 - \chi(\Delta t)|.
	\end{align}
	
	\item \textbf{Maximally mixed state:} Setting $\alpha_0 = \alpha_1 = 1/2$, $\alpha_{01} = 0$ gives
	\begin{align}
		W_Q = \frac{1}{4} |1 - \chi(\Delta t)|.
	\end{align}
	
	\item \textbf{Pure superposition state:} Setting $\alpha_0 = \alpha_1 = 1/2$, $\alpha_{01} = 1/2$ gives
	\begin{align}
		W_Q = \frac{1}{4} |1 - \chi(\Delta t)|.
	\end{align}
	This is the same as the maximally mixed case because the first measurement destroys the coherences.
	
	\item \textbf{Eigenstate:} Setting $\alpha_0 = 0$ or $\alpha_1 = 0$ gives
	\begin{align}
		W_Q = 0,
	\end{align}
	as expected.
\end{enumerate}

Several important observations follow from these results:

\begin{enumerate}
	\item The witness for projective measurements is independent of initial coherences. This is a direct consequence of the projection postulate: the first measurement collapses the state onto the measurement basis, erasing any initial quantum coherences.
	\item The witness depends only on the initial populations $\alpha_0, \alpha_1$ and the dephasing factor $\chi(\Delta t)$.
	\item The maximum value of the witness is achieved when $\alpha_0 = \alpha_1 = 1/2$ and $\chi(\Delta t) = 0$, giving
	\begin{align}
		W_Q^{\text{max}} = \frac{1}{4}.
	\end{align}
\end{enumerate}

Note that this maximum value applies specifically to dichotomic systems (i.e., qubits) with two consecutive measurements. The value of $W_Q^{\max}$ depends on the number of accessible outcomes and the number of measurements performed. For a general system with $d$ outcomes and $K$ measurements, the maximum value would in general differ from $1/4$.
\subsection{Comparison with known bounds}

To contextualize our result, it is useful to compare the maximum value of our witness with known bounds in the literature. Table~\ref{tab:comparison} summarizes the maximum values of different nonclassicality measures for a qubit.

\begin{table}[htbp]
    \centering
    \caption{Comparison of maximum values of different nonclassicality measures for a qubit.}
    \begin{tabular}{|l|c|c|}
        \hline
        \textbf{Measure} & \textbf{Maximum Value} & \textbf{Reference} \\
        \hline
        Leggett-Garg parameter \( K_3 \) & \( 3/2 \) & \cite{leggett1985quantum} \\
        Temporal Bell inequality & \( 2\sqrt{2} \) & \cite{fine1982hidden} \\
        NSIT violation & \( 1 \) & \cite{milz_kolmogorov_2017} \\
        \textbf{This work} \( W_Q^{\max} \) & \( \mathbf{1/4} \) & --- \\
        \hline
    \end{tabular}
    \label{tab:comparison}
\end{table}

The maximum value \( W_Q^{\max} = 1/4 \) is smaller than the Leggett-Garg maximum \( 3/2 \) and the NSIT violation maximum \( 1 \). This reflects the fact that our witness probes a different aspect of nonclassicality — namely, measurement unsharpness — rather than temporal correlations or macrorealism. Unlike Leggett-Garg inequalities, which can reach large values due to sequential measurements, our witness is bounded by the intrinsic unsharpness of the POVM. Thus, \( W_Q^{\max} = 1/4 \) represents the maximal contribution of measurement unsharpness to the violation of Kolmogorov consistency, making it a complementary tool to existing nonclassicality witnesses.

\begin{figure}[htbp!]
	\centering
	\includegraphics[width=0.8\textwidth]{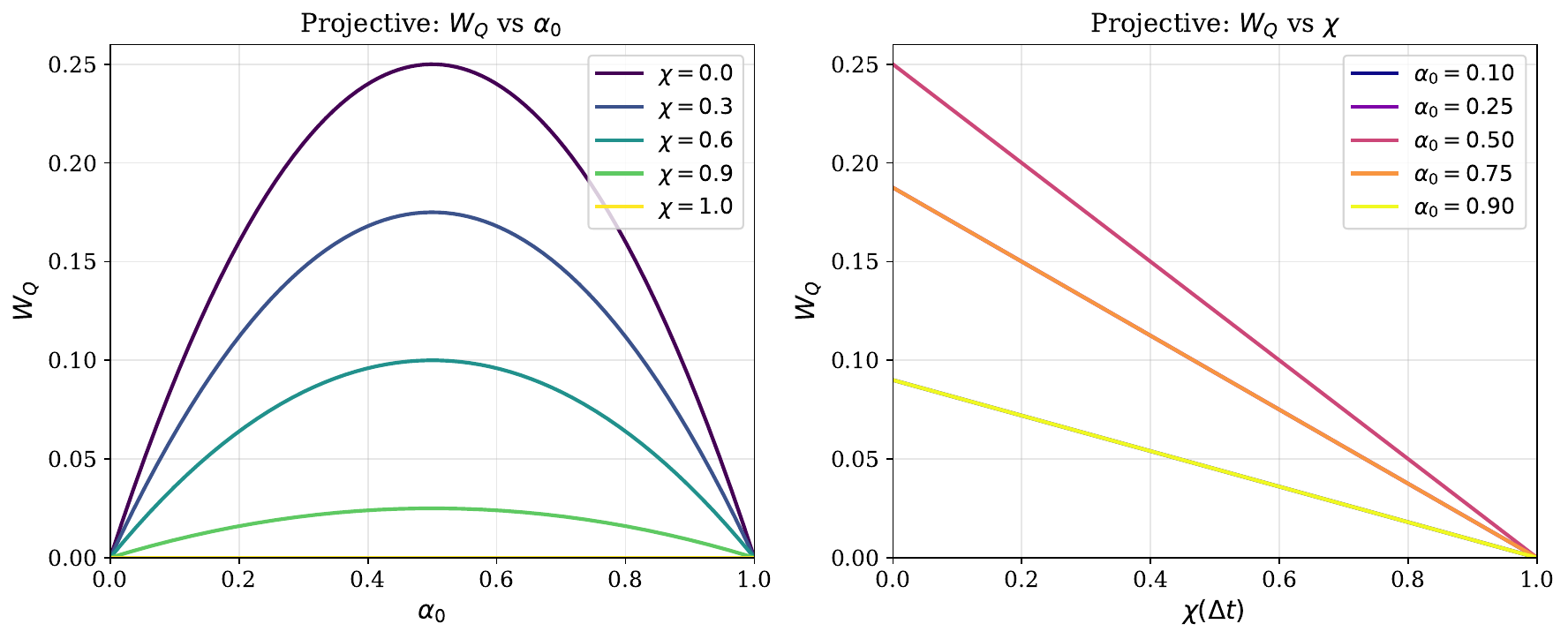}
	\caption{(Color online) Nonclassicality witness $W_Q$ for projective measurements. Panel (a) shows $W_Q$ as a function of the initial population $\alpha_0$ for different values of the dephasing factor $\chi$. The curves are parabolas with maxima at $\alpha_0 = 1/2$, corresponding to the maximally mixed initial state. As $\chi \to 1$, the witness vanishes, indicating that decoherence suppresses nonclassicality. Panel (b) shows $W_Q$ as a function of $\chi$ for different values of $\alpha_0$. The linear decrease reflects the monotonic dependence of the witness on dephasing. The maximum value $W_Q = 1/4$ is achieved at $\alpha_0 = 1/2$ and $\chi = 0$.}
	\label{fig:projective}
\end{figure}

These results are illustrated in Fig.~\ref{fig:projective}. Figure~\ref{fig:projective} illustrates the behavior of the witness for projective measurements. In panel (a), the parabolic dependence on $\alpha_0$ shows that the witness is maximized for the maximally mixed state, where the uncertainty in the initial population is largest. The suppression of the witness as $\chi$ increases reflects the effect of dephasing: stronger decoherence reduces the nonclassicality detectable by the witness. Panel (b) confirms this linear dependence on $\chi$ for different population imbalances.

\section{Generalized measurements}
\label{sec:povm}

In the previous section, we restricted our attention to projective measurements in the computational basis. However, in many experimental scenarios, measurements are more naturally described by positive operator-valued measures (POVMs). In this section, we generalize our framework to accommodate POVMs and derive the corresponding nonclassicality witness.

\begin{definition}[POVM]
A positive operator-valued measure (POVM) is a set of operators $\{E_x\}_{x \in \mathcal{X}}$ such that (see Refs.~\cite{QPQI-Book, Preskill-2019})
\begin{gather}
	E_x \geq 0 \quad \forall x \in \mathcal{X}, \qquad \sum_{x \in \mathcal{X}} E_x = \mathbb{I},
\end{gather}
where $\mathcal{X}$ is a finite set of measurement outcomes. The probability of obtaining outcome $x$ when measuring a system in state $\rho$ is given by the Born rule:
\begin{gather}
	P(x) = \Tr[E_x \rho].
\end{gather}
\end{definition}

For a qubit system, the most general POVM with two outcomes $\{0,1\}$ can be parametrized as
\begin{align}
	\label{eqn:povm_qubit}
	E_0 &= \frac{1}{2}\left( \mathbb{I} + \vec{a} \cdot \vec{\sigma} \right), \\
	E_1 &= \frac{1}{2}\left( \mathbb{I} - \vec{a} \cdot \vec{\sigma} \right),
\end{align}
where $\vec{a} \in \mathbb{R}^3$ with $|\vec{a}| \leq 1$, and $\vec{\sigma} = (\sigma_x, \sigma_y, \sigma_z)$ is the vector of Pauli matrices. The condition $|\vec{a}| \leq 1$ ensures that $E_0, E_1 \geq 0$. Projective measurements correspond to the special case $|\vec{a}| = 1$.

In the sequential measurement scheme, we now replace the projective measurements with POVMs. The joint probability of obtaining outcome $x_1$ at time $t_1$ and outcome $x_2$ at time $t_2$ is given by
\begin{align}
	\label{eqn:joint_prob_povm}
	P(x_2,t_2; x_1,t_1) &= \Tr\left[ E_{x_2} \Lambda_{t_2,t_1}\left( \sqrt{E_{x_1}} \rho_0 \sqrt{E_{x_1}} \right) \right],
\end{align}
where we have used the measurement update rule for POVMs, and $\rho_0$ is the initial state of the system given in Eq.~\eqref{eqn:state_general}.

In the simplest case of a qubit initialized in the diagonal state (i.e., $\alpha_{01} = 0$) and undergoing pure dephasing dynamics, the nonclassicality witness takes the form
\begin{align}
	\label{eqn:witness_povm}
	W_Q &= \frac{1}{4} \left| 1 - \chi(\Delta t) \right| \left| 1 - a_z^2 \right| \left| 1 - \tanh^2(\beta/2) \right|,
\end{align}
where $a_z$ is the $z$-component of the POVM vector $\vec{a}$, and $\beta$ parametrizes the initial state populations via $\alpha_0 = e^{\beta}/(e^{\beta} + e^{-\beta})$.

For the general initial state including coherences $\alpha_{01}$, the witness becomes
\begin{align}
	\label{eqn:witness_povm_general}
	W_Q &= \frac{1}{4} \left| 1 - \chi(\Delta t) \right| \left| 1 - a_z^2 \right| \left| 1 - \tanh^2(\beta/2) \right| + \mathcal{O}(\alpha_{01}),
\end{align}
where the terms involving $\alpha_{01}$ depend on the specific POVM parameters and vanish for projective measurements.

The term $\mathcal{O}(\alpha_{01})$ denotes first-order corrections in the initial coherence $\alpha_{01}$. These corrections vanish exactly for diagonal initial states ($\alpha_{01} = 0$), which is the focus of our analysis. They also vanish in the projective limit $|a_z| = 1$, where the POVM reduces to a projective measurement. For small coherences ($|\alpha_{01}| \ll 1$), these corrections are negligible and do not affect the qualitative behavior of the witness. In the weak-measurement regime ($|a_z| \ll 1$), the corrections are further suppressed by the measurement strength.

This result generalizes the projective case. Indeed, for projective measurements $|a_z| = 1$, the witness vanishes identically, recovering the fact that projective measurements on a diagonal state do not reveal nonclassicality. For $|a_z| < 1$, however, the witness can be nonzero, provided that $\chi(\Delta t) \neq 1$ and the initial state is not an eigenstate ($\tanh^2(\beta/2) \neq 1$).

The maximum value of the witness in the POVM case is achieved when $a_z = 0$, $\chi(\Delta t) = 0$, and $\alpha_0 = \alpha_1 = 1/2$, giving
\begin{align}
	\label{eqn:witness_povm_max}
	W_Q^{\text{max}} &= \frac{1}{4}.
\end{align}

Remarkably, this maximum is the same as in the projective case, but it is now achieved for a wider range of measurements, including non-projective POVMs. This demonstrates that the nonclassicality witness is robust and can be applied in experimental settings where only generalized measurements are available.
\subsection{Measurement nonclassicality vs state nonclassicality}

The results above reveal a fundamental distinction between our witness and standard nonclassicality measures. Unlike conventional witnesses, which probe the nonclassicality of the quantum state and vanish for classical states, our POVM-based witness does not vanish even when the initial state is diagonal in the measurement basis. This indicates that the nonclassicality detected here is not a property of the state, but rather a property of the measurement itself.

For projective measurements, the witness vanishes identically for diagonal initial states ($W_Q = 0$), as shown in Sec.~\ref{sec:results}. In contrast, POVMs with $|a_z| < 1$ yield $W_Q > 0$ for the same classical initial states. This demonstrates that the nonclassicality revealed by POVMs is genuinely new — it is not already detectable by projective measurements, nor can it be attributed to the state alone. Rather, it arises from the inherent unsharpness of the measurement, which allows coherences to persist and contribute to the violation of Kolmogorov consistency.

Thus, our witness provides an operational tool for quantifying measurement nonclassicality — a concept that is complementary to state-based nonclassicality and inaccessible to projective measurements. This addresses a gap in the literature, where most witnesses focus on states or temporal correlations, and highlights the resource-theoretic role of measurement unsharpness in quantum information processing.

\begin{figure}[htbp!]
	\centering
	\includegraphics[width=0.8\textwidth]{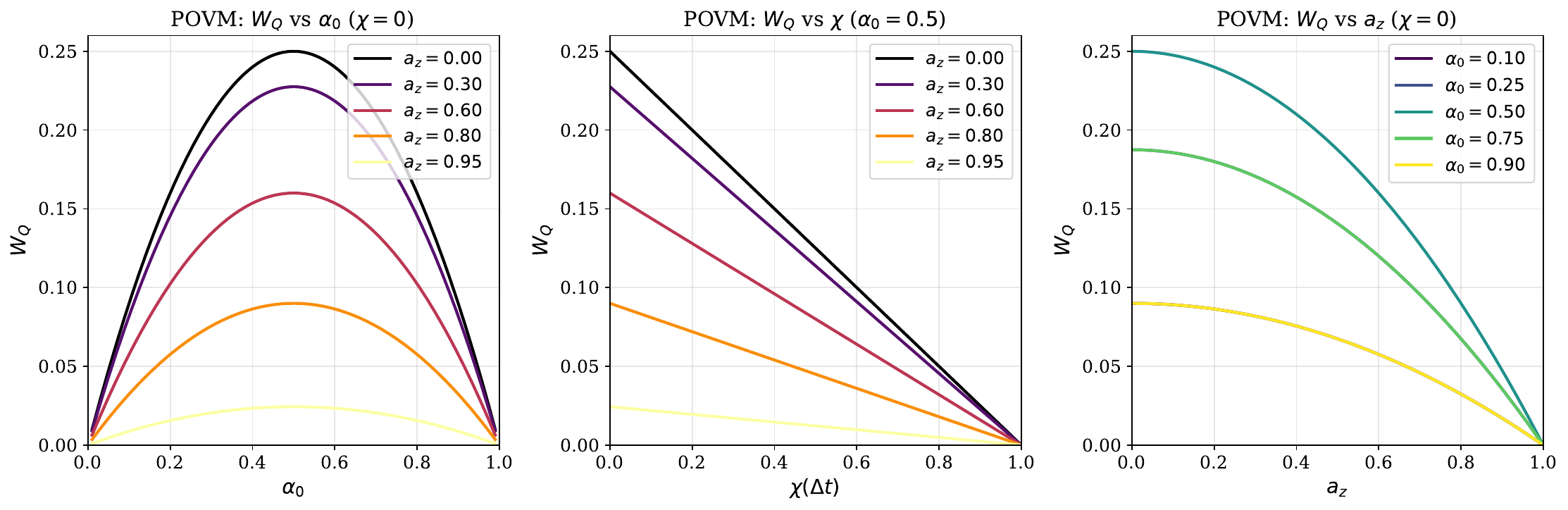}
\caption{(Color online) Nonclassicality witness $W_Q$ for POVM measurements. Panel (a) shows $W_Q$ as a function of the initial population $\alpha_0$ for different values of the POVM parameter $a_z$, with $\chi = 0$. As $a_z$ decreases (i.e., the measurement becomes more unsharp), the witness increases, reaching its maximum for $a_z = 0$. Panel (b) shows $W_Q$ as a function of $\chi$ for different values of $a_z$, with $\alpha_0 = 0.5$, confirming that the witness is maximized for unbiased POVMs. Panel (c) shows $W_Q$ as a function of $a_z$ for different values of $\alpha_0$, with $\chi = 0$, highlighting the projective limit $a_z \to 1$, where the witness vanishes.}	\label{fig:povm}
\end{figure}

\begin{figure}[htbp!]
	\centering
	\includegraphics[width=0.6\textwidth]{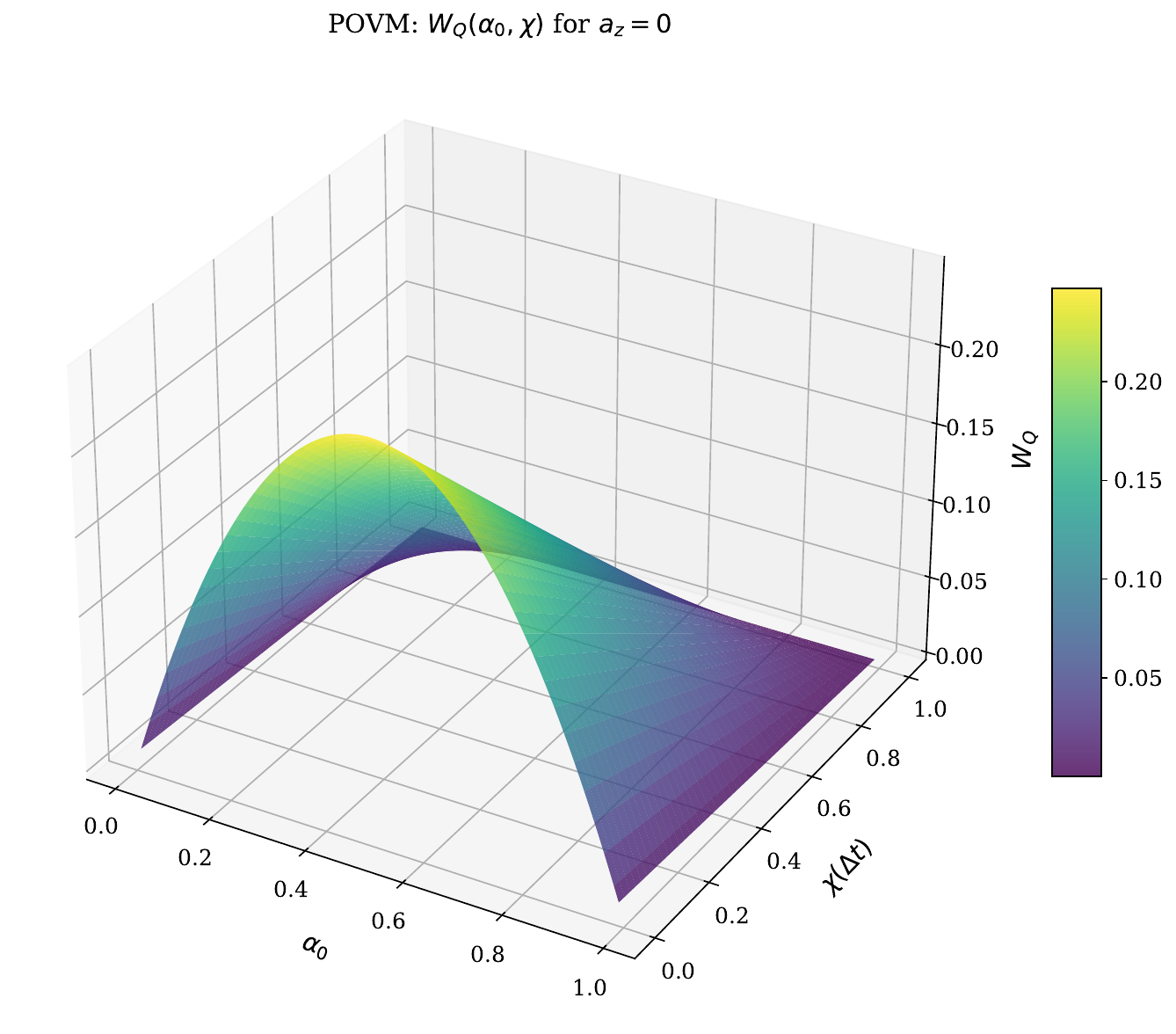}
	\caption{(Color online) Three-dimensional plot of the nonclassicality witness $W_Q(\alpha_0, \chi)$ for a qubit measured with an unbiased POVM ($a_z = 0$). The surface shows a clear peak at $\alpha_0 = 1/2$ and $\chi = 0$, corresponding to the maximally mixed initial state and the absence of dephasing. The maximum value is $W_Q^{\max} = 1/4$. The smooth decay to zero along both axes illustrates the trade-off between initial state purity and environmental decoherence.}
	\label{fig:3d}
\end{figure}

\begin{figure}[htbp!]
	\centering
	\includegraphics[width=0.8\textwidth]{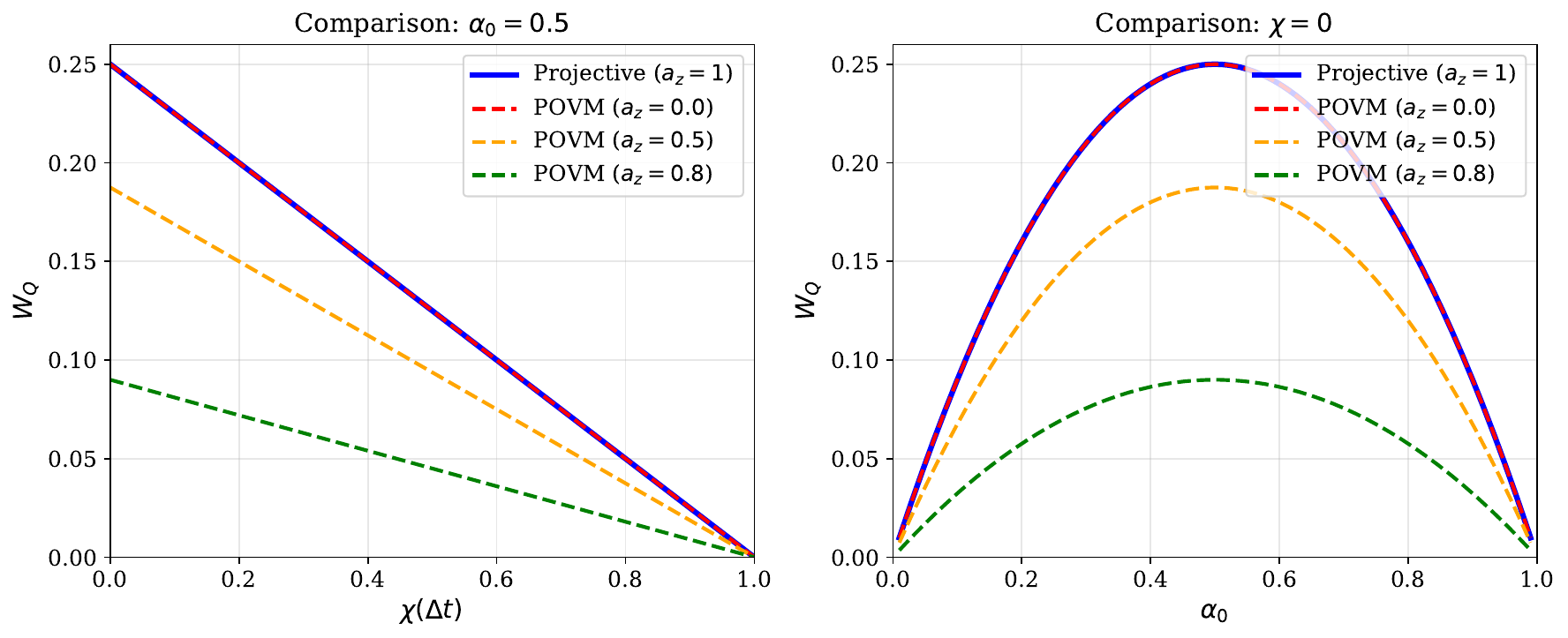}
	\caption{(Color online) Comparison between projective measurements ($a_z = 1$, solid blue line) and POVMs ($a_z < 1$, dashed lines). Panel (a) shows $W_Q$ as a function of the dephasing factor $\chi$ for fixed $\alpha_0 = 0.5$. Projective measurements yield $W_Q = 0$ identically, while POVMs yield nonzero values that increase as $a_z$ decreases. Panel (b) shows $W_Q$ as a function of $\alpha_0$ for fixed $\chi = 0$, confirming the same contrast: projective measurements give zero witness, whereas POVMs reveal nonclassicality for all population imbalances. The POVM curves approach the projective limit as $a_z \to 1$. This comparison clearly demonstrates that POVMs are essential for witnessing nonclassicality in diagonal initial states.}
	\label{fig:comparison}
\end{figure}

These results are illustrated in Figs.~\ref{fig:povm}, \ref{fig:3d}, and \ref{fig:comparison}. 

Figure~\ref{fig:povm} generalizes the analysis to POVMs. Panel (a) demonstrates that the witness increases as the POVM becomes more unsharp (i.e., as $a_z$ decreases), reaching its maximum for the unbiased case $a_z = 0$. Panel (b) shows that this behavior persists across different values of the dephasing factor $\chi$, with the witness vanishing only in the projective limit. Panel (c) explicitly shows the dependence on $a_z$, confirming that the witness is zero for projective measurements ($|a_z| = 1$) and maximal for unbiased POVMs ($a_z = 0$).

Figure~\ref{fig:3d} provides a comprehensive view of the witness landscape for an unbiased POVM. The surface plot shows a sharp peak at $\alpha_0 = 1/2$ and $\chi = 0$, where the witness reaches its maximum value $1/4$. The smooth decay to zero along the $\alpha_0$ and $\chi$ axes reflects the loss of nonclassicality as the initial state becomes pure or as decoherence increases.

Figure~\ref{fig:comparison} directly contrasts projective and POVM measurements. In panel (a), the projective case (solid blue line) gives $W_Q = 0$ for all values of $\chi$, while POVMs (dashed lines) yield positive values that increase as the measurement becomes more unsharp (smaller $a_z$). Panel (b) shows the same behavior as a function of $\alpha_0$, confirming that projective measurements fail to detect nonclassicality for diagonal states, whereas POVMs consistently reveal it. This comparison highlights the central message of our work: measurement unsharpness is a resource for witnessing nonclassicality.

\section{Discussion and Conclusion}
\label{sec:discussion}

We have introduced an operational measure of nonclassicality for qubit systems based on the violation of Kolmogorov consistency conditions in sequential measurements. Our main results can be summarized as follows.

First, we established that for projective measurements on any initial state, the nonclassicality witness $W_Q = \alpha_0 \alpha_1 |1 - \chi(\Delta t)|$ is independent of initial coherences. This is because the first projective measurement destroys all coherences, so the witness only probes coherences generated by the dynamics. This result confirms that projective measurements alone cannot reveal nonclassicality arising from initial coherences—any observed effects are purely dynamical.

Second, by generalizing to POVMs, we found that non-projective measurements can reveal nonclassicality even for diagonal initial states. The witness takes the form $W_Q = \frac{1}{4} |1 - \chi(\Delta t)| |1 - a_z^2| |1 - \tanh^2(\beta/2)|$ for diagonal states, which vanishes only in the projective limit $|a_z| = 1$. For general initial states with coherences, additional contributions to $W_Q$ appear, highlighting the role of measurement incompatibility in revealing both initial and dynamically generated quantum features.

Third, the maximum value of the witness is $W_Q^{\text{max}} = 1/4$ in all cases, achieved when $\alpha_0 = \alpha_1 = 1/2$, $\chi(\Delta t) = 0$, and (for POVMs) $a_z = 0$. This universality suggests that the measure captures a fundamental aspect of nonclassicality that is independent of the measurement implementation.

Fourth, we have provided a microscopic derivation of the dephasing channel from a system-bath Hamiltonian, establishing the physical foundation of our phenomenological model. This derivation shows that the dephasing factor $\chi(\Delta t)$ encodes the spectral density, temperature, and coupling strength of the bath, and that the Markovian limit recovers the simple exponential decay commonly used in quantum information theory.

Our framework connects several important concepts in quantum foundations: Kolmogorov consistency, Leggett-Garg inequalities, NCGD maps, POVMs, and open quantum systems. The witness we introduced provides an experimentally accessible tool for detecting nonclassicality in qubit systems, with potential applications in quantum technology certification and quantum resource theory.

Future work could extend this framework to higher-dimensional systems, non-Markovian dynamics (see Ref.~\cite{Taranto-2020}), and more general measurement schemes. Additionally, connecting our measure to other nonclassicality quantifiers, such as quantum discord or entanglement, would provide a more complete picture of the quantum-classical boundary.

In conclusion, we have shown that POVMs provide a powerful tool for revealing nonclassicality in qubit systems, even when the initial state is diagonal in the measurement basis. Our results underscore the importance of generalized measurements in quantum information processing and provide a practical witness for experimental quantum technology applications.

\section*{Acknowledgments}
We acknowledge helpful discussions with colleagues. A.S.K. thanks the Wilczek Quantum Center, Shanghai Jiao Tong University, for hospitality during the early stages of this work.

\appendix

\section{Spin-boson model for pure dephasing}
\label{app:spinboson}

\subsection{System-bath Hamiltonian}

We consider a qubit coupled to a bosonic bath. The total Hamiltonian is

\begin{align}
    H = \frac{\omega_0}{2} \sigma_z + \sum_k \omega_k b_k^\dagger b_k + \sigma_z \sum_k g_k (b_k + b_k^\dagger),
\end{align}

where $\omega_0$ is the qubit energy splitting, $\omega_k$ are the bath frequencies, $b_k$ ($b_k^\dagger$) are the bosonic annihilation (creation) operators, and $g_k$ are the system-bath coupling constants. The bath is characterized by its spectral density

\begin{align}
    J(\omega) = \sum_k g_k^2 \delta(\omega - \omega_k).
\end{align}

\subsection{Master equation}

Assuming the bath is initially in a thermal state $\rho_B = e^{-\beta H_B}/\Tr[e^{-\beta H_B}]$ at inverse temperature $\beta = 1/(k_B T)$, and working in the interaction picture, the reduced dynamics of the qubit is described by the master equation

\begin{align}
    \dot{\rho}_S(t) = -\frac{i}{\hbar}[H_S, \rho_S(t)] - \gamma(t) [\sigma_z, [\sigma_z, \rho_S(t)]],
\end{align}

where the time-dependent dephasing rate is

\begin{align}
    \gamma(t) = \int_0^t ds \, \nu(s),
\end{align}

and $\nu(s)$ is the bath correlation function

\begin{align}
    \nu(s) = \int_0^\infty d\omega \, J(\omega) \coth\left(\frac{\beta \hbar \omega}{2}\right) \cos(\omega s).
\end{align}

\subsection{Solution of the master equation}

The solution yields the dephasing channel

\begin{align}
    \Lambda_{t_2,t_1}(\rho) = \sum_{i,j=0}^1 \chi_{ij}(t_2-t_1) \ket{i}\bra{i} \rho \ket{j}\bra{j},
\end{align}

with $\chi_{01}(t) = \chi_{10}^*(t) = \chi(t)$, and the dephasing factor given by

\begin{align}
    \chi(t) = \exp\left[-\int_0^\infty d\omega \, J(\omega) \coth\left(\frac{\beta \hbar \omega}{2}\right) \frac{1 - \cos(\omega t)}{\omega^2}\right].
\end{align}

\subsection{Special cases}

For an Ohmic bath with spectral density $J(\omega) = \eta \omega e^{-\omega/\omega_c}$, the zero-temperature limit gives

\begin{align}
    \chi(t) = \exp\left[-\frac{\eta}{2} \left(1 - e^{-\omega_c t}\right)\right].
\end{align}

In the Markovian limit, where the bath correlation time is much shorter than the system timescale, this reduces to

\begin{align}
    \chi(t) = e^{-\gamma t},
\end{align}

with constant dephasing rate $\gamma$.

\subsection{Connection to the witness}

This microscopic model provides the physical foundation for the dephasing factor $\chi(\Delta t)$ used in the main text. It allows us to express the witness in terms of experimentally tunable parameters such as temperature, coupling strength, and cutoff frequency.

\bibliographystyle{apsrev4-2}
\bibliography{ref}

@article{piron_ideal_1981,
  author = {Piron, C.},
  title = {Foundations of Quantum Physics},
  journal = {Book},
  publisher = {Benjamin},
  address = {Reading, MA},
  year = {1981}
}

@article{BreuerEA2016,
  author = {Breuer, H.-P. and Laine, E.-M. and Piilo, J. and Vacchini, B.},
  title = {Colloquium: Non-Markovian dynamics in open quantum systems},
  journal = {Reviews of Modern Physics},
  volume = {88},
  pages = {021002},
  year = {2016},
  doi = {10.1103/RevModPhys.88.021002}
}

@article{milz_kolmogorov_2017,
  author = {Milz, S. and Pollock, F. A. and Modi, K.},
  title = {Reconstructing non-Markovian quantum dynamics},
  journal = {Physical Review A},
  volume = {96},
  pages = {052115},
  year = {2017},
  doi = {10.1103/PhysRevA.96.052115}
}

@article{fine1982hidden,
  author = {Fine, A.},
  title = {Hidden Variables, Joint Probability, and the Bell Inequalities},
  journal = {Physical Review Letters},
  volume = {48},
  pages = {291},
  year = {1982},
  doi = {10.1103/PhysRevLett.48.291}
}

@article{leggett1985quantum,
  author = {Leggett, A. J. and Garg, A.},
  title = {Quantum mechanics versus macroscopic realism: Is the flux there when nobody looks?},
  journal = {Physical Review Letters},
  volume = {54},
  pages = {857},
  year = {1985},
  doi = {10.1103/PhysRevLett.54.857}
}

@book{kolmogorov_foundations_1956,
  author = {Kolmogorov, A. N.},
  title = {Foundations of the Theory of Probability},
  publisher = {Chelsea},
  address = {New York},
  year = {1956}
}

@book{feller_introduction_1968,
  author = {Feller, W.},
  title = {An Introduction to Probability Theory and Its Applications},
  volume = {1},
  publisher = {Wiley},
  address = {New York},
  year = {1968}
}

@book{breuer_theory_2007,
  author = {Breuer, H.-P. and Petruccione, F.},
  title = {The Theory of Open Quantum Systems},
  publisher = {Oxford University Press},
  address = {Oxford},
  year = {2007}
}

@book{tao_introduction_2011,
  author = {Tao, T.},
  title = {An Introduction to Measure Theory},
  publisher = {American Mathematical Society},
  address = {Providence},
  year = {2011}
}

@article{DelSanto-2025,
  author = {Del Santo, F. and Gisin, N.},
  title = {Which features of quantum physics are not fundamentally quantum but are due to indeterminism?},
  journal = {Quantum},
  volume = {9},
  pages = {1686},
  year = {2025}
}

@article{Smirne-2019,
  author = {Smirne, A. and Egloff, D. and Díaz, M. G. and Plenio, M. B. and Huelga, S. F.},
  title = {Coherence and non-classicality of quantum Markov processes},
  journal = {Quantum Science and Technology},
  volume = {4},
  pages = {01LT01},
  year = {2019},
  doi = {10.1088/2058-9565/aae9c7}
}

@article{Milz-2020,
  author = {Milz, S. and Egloff, D. and Taranto, P. and Theurer, T. and Plenio, M. B. and Smirne, A. and Huelga, S. F.},
  title = {When Is a Non-Markovian Quantum Process Classical?},
  journal = {Physical Review X},
  volume = {10},
  pages = {041049},
  year = {2020},
  doi = {10.1103/PhysRevX.10.041049}
}

@article{Taranto-2020,
  author = {Taranto, P.},
  title = {Memory Effects in Quantum Processes},
  journal = {International Journal of Quantum Information},
  volume = {18},
  pages = {1941002},
  year = {2020},
  doi = {10.1142/S0219749919410029}
}

@book{QPQI-Book,
  author = {Nielsen, M. A. and Chuang, I. L.},
  title = {Quantum Computation and Quantum Information},
  publisher = {Cambridge University Press},
  year = {2010}
}

@article{Preskill-2019,
  author = {Preskill, J.},
  title = {Lecture Notes for Physics 219: Quantum Computation},
  journal = {Lecture Notes},
  chapter = {3},
  year = {2019}
}

@article{Sakuldee-2022a,
  author = {Sakuldee, Fattah and Cywi{\'n}ski, {\L}ukasz},
  title = {Statistics of projective measurement on a quantum probe as a witness of noncommutativity of algebra of a probed system},
  journal = {Quantum Information Processing},
  volume = {21},
  number = {7},
  pages = {244},
  year = {2022}
}

@article{Sakuldee-2022b,
  author = {Sakuldee, Fattah and Taranto, Philip and Milz, Simon},
  title = {Connecting commutativity and classicality for multitime quantum processes},
  journal = {Physical Review A},
  volume = {106},
  pages = {022416},
  year = {2022},
  doi = {10.1103/PhysRevA.106.022416}
}

\end{document}